\documentclass{article}
\usepackage{amsmath}
\usepackage{amstext}
\usepackage{amsfonts}
\usepackage{amssymb,amscd}
\usepackage{amsopn}
\usepackage{graphicx,color}
\usepackage{cite}
\usepackage{bm,bbm}
\usepackage{epsfig} %,epic}

\begin{document}

%\newpage  
\vspace{0.3cm}

\begin{center}

{\Large Bogdan Mielnik, Geometry and Quanta}

\vspace{0.3cm}

{\large 
Ingemar Bengtsson\footnote{Stockholms Universitet,
 Fysikum,
ORCID: 0000-0002-4203-3180, 

email: ingemar@fysik.su.se}
     and
 Karol \.Zyczkowski\footnote{Institute of Theoretical Physics, 
 Jagiellonian University, Cracow and 
 Center for Theoretical Physics, Polish Academy of Sciences, Warsaw;
 ORCID: 0000-0002-0653-3639,
 
  email: karol@tatry.if.uj.edu.pl}
 }

%May 3,  2022

\end{center}
\vspace{13mm}
 {\bf Abstract}. 
We review selected achievements of the late Bogdan Mielnik in the field 
of theoretical physics, with an emphasis on his attempts to go beyond quantum mechanics. 
Some of his original views on the problems of contemporary society
and organization of science are also recalled.
\vspace{5mm}

{\sl keywords: Quantum Theory, Foundations of Physics, Bogdan Mielnik}

\vspace{5mm}

{\bf Bogdan Mielnik} (1936-2019) graduated in theoretical physics in 1958
 at the University of Warsaw, which at the time was one of the select few 
 places where the renaissance of general 
 relativity as a central field of physics was being prepared. Invited by his supervisor 
 Jerzy Pleba{\'n}ski, Bogdan went to Mexico and 
 submitted his PhD thesis  \cite{Mie64} on October 22, 1964.
 He was the very first Ph.D. graduate of the Physics Department  at the
 Center for Research and Advanced Studies of the National Polytechnic Institute
 {\sl Cinvestav} founded in Mexico City in 1961. He returned to Poland in 1965. In 1975 he 
 made a memorable visit to Stockholm, one of the quiet places where the theory 
 of open quantum systems saw the light of day. He 
 went again to Mexico in 1981 and afterwards shared his time between the
 University of Warsaw and Cinvestav. Eventually he described some of his 
 adventures in the Pleba{\'n}ski Festschrift \cite{Plebanski}. 
 
\begin{figure}[h]%disk
%     \hskip 1.5cm
     \center{
	\includegraphics[angle=0,width=0.30\columnwidth]{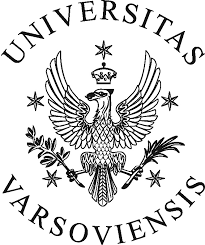}
	\hskip 1.9cm
	\includegraphics[angle=0,width=0.30\columnwidth]{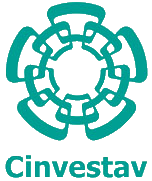}}
	\caption{The entire scientifc career of Bogdan Mielnik was strongly  linked to the
	University of Warsaw and Cinvestav in Mexico City.}
	\label{fig:1}
\end{figure}
 \medskip
 
  A central theme of Bogdan's work was to understand why quantum mechanics takes the shape it 
  does, and whether it needs to be changed (perhaps when taking gravity into account). When 
  he started out quantum logic gave its answers to the first question, and made it seem as 
  if no changes in the formalism were possible. Bogdan---inspired by G\"unther Ludwig and 
  by Rudolf Haag, if we read his reference lists correctly---took a more 
  geometrical view of the axioms. Instead of a lattice of propositions his starting point 
  was the more flexible lattice of faces of convex bodies. In the late sixties 
  he published two notable papers \cite{Mie68,Mie69}  in Commun. Math. Phys.,
  in which he explained how the geometry of the set of states of a physical system is 
  determined by the transition probabilities associated to idealized experiments. If we 
  impose suitable `crystalline symmetries' on the convex body (the {\sl statistical figure} 
  in Bogdan's terminology) the lattice of faces can be identified with the lattice of 
  subpaces in a Hilbert space, and hence quantum logic is recovered. 
  
 Here we must admit that when (many years later) we wrote a book on the geometry of 
 quantum theory we borrowed its title \cite{BZ06} from one of these papers. 
  Fascinated by the originality of his thoughts we also opened our book with 
  a quotation from another paper of his \cite{Mie81}:
  
  \medskip
\noindent {\sl  What picture does one see, looking at a physical theory from a distance, so that the details
disappear? Since quantum mechanics is a statistical theory, the most universal picture which
remains after the details are forgotten is that of a convex set.}
\medskip 

\noindent Bogdan's pioneering ideas were appreciated later on, when the development 
  of the field of quantum information science made the space of quantum states into a stage on which quantum algorithms can be played. 

\begin{figure}[h]
    \center{
	\includegraphics[angle=0,width=0.85\columnwidth]{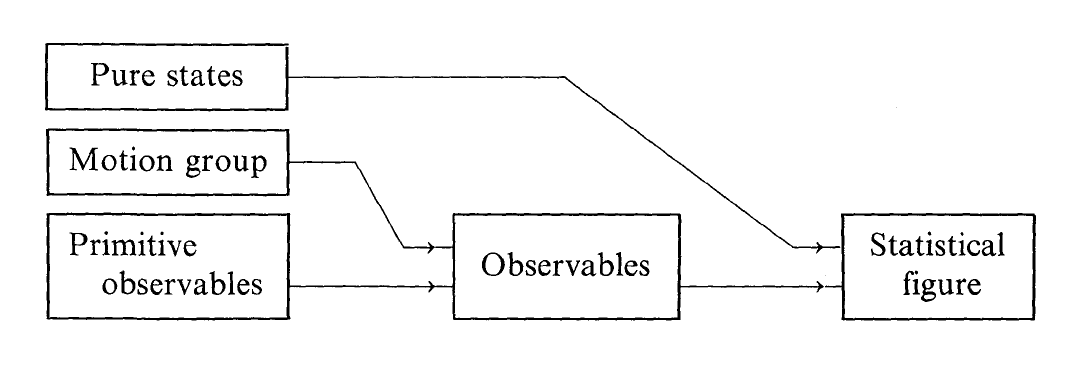}}
	\caption{Sketch of the  intrinsic structure of quantum mechanics presented in 
  the  paper of Bogdan on generalized quantum theories \cite{Mie74}.}
	\label{fig:2}
\end{figure}

In particular Bogdan took an original view of the concept of the state of a system. 
For him the physical system is always open, and subject to the influences of the universe 
in which it is placed. It is this {\sl mobility} that defines the system. This means 
that the one-parameter group transformations generated by a 
definite Hamiltonian loses its central status. Instead we have to ask for 
the {\sl motion group}, the set of all transformations that the universe may cause 
the system to undergo. This leads to the {\sl dynamical manipulation problem} that 
Bogdan initiated in Stockholm, this time inspired by the works of Willis Lamb and Elihu Lubkin. 
This problem has a rich technical content in the framework of orthodox quantum mechanics, 
and it was to occupy much of his attention in later years (although the idea to challenge 
orthodoxy was still with him, as one can see in his grand review of factorization and supersymmetric 
quantum mechanics \cite{Rosas}). 

With a persistent enthusiasm Bogdan analyzed the interrelations between quantum theory and 
general relativity. He stressed the flexibility introduced by the latter theory when it 
went from a fixed spacetime to a dynamically determined one, and he felt---not surprisingly 
given his background---that joining it to the rigid structure of quantum mechanics may 
violate the {\sl innate aesthetics of general relativity} \cite{Mie76}. 

In the end no satisfactory alternative presented itself. And over the past fifty years 
some very important developments have taken place: Quantum theory has been tested 
in the laboratory in a wide range of qualitatively new circumstances, and it has never 
been found wanting. But one of Bogdan's papers \cite{Mie81} starts with a quotation from 
Stanis{\l}aw Lem, worth repeating here: {\sl It is well known that dragons do not exist. 
But each one does it in a different way.} Bogdan presumably quoted from memory, because his 
version is snappier than the original. These words can be considered as a prophetic 
prediction of the enormous scientific interest in generalized probabilistic theories 
developed in the current century, and intended to go beyond standard quantum mechanics. The 
contributions in the book edited by Chiribella and Spekkens  \cite{CS}
can be recommended for an 
update. We mean no offence if we add that for charm, originality, and seriousness, 
Bogdan's papers still stand supreme. 

And concerning quantum gravity, there is as yet no need to make changes to the conclusion 
of his 1974 paper \cite{Mie74}: {\sl 
The incompleteness of the present day science at this point is, perhaps, one more
reason why the scheme of quantum mechanics should not be prematurely 
closed}. 

  \medskip
  
    \medskip
  
In the late spring of 2000 Ingemar was visiting Poland during a time when Bogdan 
was temporarily working in Warsaw. We met him 
 in the small bar  at the old building of the Faculty of Physics 
 at Ho{\.z}a street, and our discussion turned out to be 
 inspiring and unforgettable. We recall the details after 20 years.
  Sipping barszcz (beet root soup) we learned more about the key
   ideas of Bogdan on what he referred to as the game of quantum mechanical pick-a-stick 
   (or `jack-straws'---none of us knew the English term). 
   Others know this game as the problem of quantum control \cite{FM94}. We also learned 
   about his ideas concerning nonlinear quantum mechanics \cite{Mie01} and, while  
   savouring pierogi, we could appreciate his novel approach to the Floquet theory \cite{DCM97,CCM06}   of time dependent quantum systems. Bogdan's 
    story was told in such a convincing and  emotional way
 that we felt that we should put aside
 all current projects and start working on this very topic.

   \smallskip

  A decade later all three of us met again 
   in Bia{\l}owie{\.z}a at the 2011 workshop 
   {\sl Geometric Methods in Physics}, 
  organized by Anatol Odzijewicz and his colleagues from Bia{\l}ystok 
  and Mexico  to celebrate the 75-th birthday of Bogdan. It was a remarkable
   experience to attend several talks explaining the contribution of Bogdan to  
   our understanding of quanta, and to hear Bogdan himself 
   describe the state of the art of his programme to explore the role of convex geometry 
   at the root of physics \cite{Mie12}. 
    Furthermore, we had  a unique chance to meet a crowd of Bogdan's friends
    and former students from Mexico, and to see how much they admire 
    him and his achievements. 
 
 Getting to know the Mexican friends of Bogdan in Bia{\l}owie{\.z}a Karol was also
 admitted to the clan and invited to attend 
 a {\sl Quantum Fest} event in Mexico City. 
 During this conference
 Bogdan presented his view on Non-Inertial Quantization \cite{CCM16}.
 It was a remarkable experience
 to watch Bodgan in action in his Spanish-speaking environment.
 Each time he appeared in the Institute or entered a conference
 room or even a dining room 
 he immediately attracted attention, as numerous people 
 wanted to hear what he had to say. It was clear to us that  
    Bogdan Mielnik has played a key role in the development 
    of theoretical physics in Mexico.\footnote{It is worth mentioning that 
    Bogdan has an extensive wikipedia webpage in the Spanish version, but up till now, he has none in Polish nor in English.}
   A brief review of his 
    activity there was presented  by Fernandez \cite{Fer13}.
    
\begin{figure}[h]
%    \hskip 0.9cm
\center{
\includegraphics[angle=0,width=0.40\columnwidth]{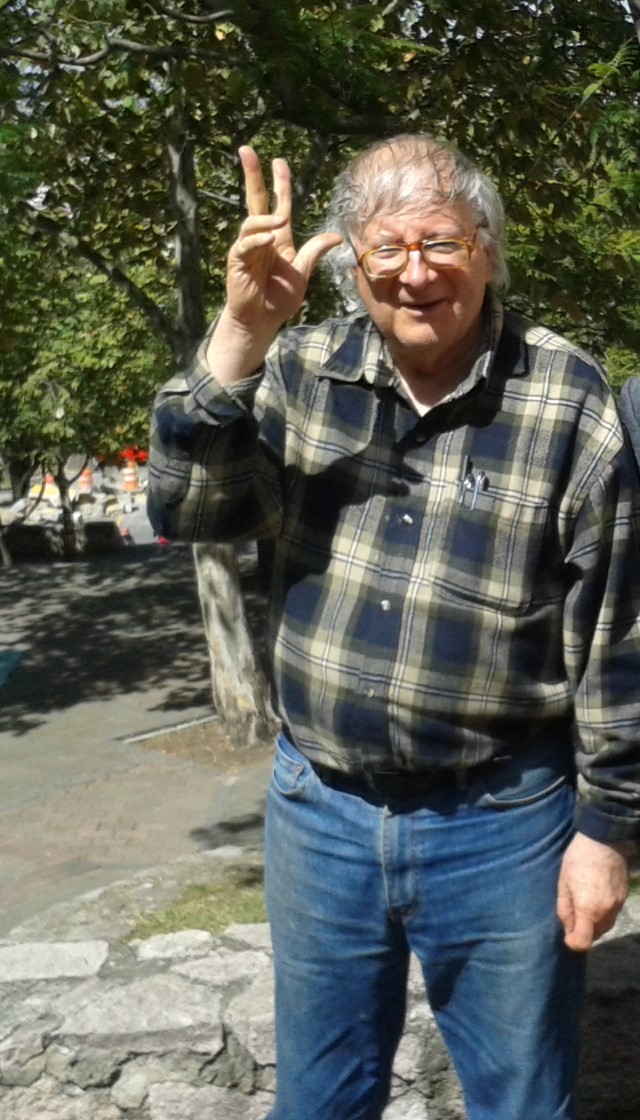}}
\caption{Bogdan Mielnik during the conference
  {\sl Quantum Fest}, 
  Tecnol{\'o}gico de Monterrey, Campus Estado de M{\'e}xico,
  October 2015.\ {\sl  Fot. Marco Enriquez}}

	\label{fig:3}
\end{figure}
 
Visiting Mexico City for the very first time is a kind of an adventure
even for an accomplished globetrotter. One can be  amazed even
looking at the trafic jams in broad streets with several lanes,
 which are not resolved by highways constructed in parallel, 
the upper lanes just above the lower lanes. For any visitor
driving in this city looks as a real challenge
and orientation without any navigation devices seems impossible.

Bogdan knew the city very well and he
explained to us his way of using cabs,
he mastered far before the age of smartphones.  
 {\sl You do not tell the driver your final destination.
 You just look around and keep saying:} {\it
 La derecha, por favor. Gire a la izquierda aqu{\' i},}
 {\sl until you reach your destination. This is simple. And is really safe.} 
 
 It looks like the life in Mexico City is not very easy, unless one gets used to it. 
 As Bogdan did, as he really enjoyed his stay in Mexico and his work 
 in this culture.
  
   \medskip  
   
We had a privilege to discuss with Bogdan numerous issues 
far beyond current problems of quantum theory. In particular we were
interested to learn his original opinions on contemporary
society,  organization of science \cite{Mie13} and
recent  economic  problems \cite{Mie15}. 
To give reader a glimpse  of his thoughts in this direction
we present in the Appendix some quotations from the  
article on bureaucracy in science \cite{Mie13}.

\bigskip
\bigskip    

%\appendix
{\bf \large Appendix.  Bogdan's thoughts on the Bureaucratic World}
\medskip

\noindent Selected quotations from the paper of Bogdan \cite{Mie13}
are presented below.

\bigskip

\centerline{* * * } 

\bigskip
{\sl

\noindent Esteemed Colleagues: The remarks below concern a certain lack of equilibrium in the present day legislation, affecting 
the life and science, with rather adverse consequences for our work. 

\medskip 

The damage to science consists not
only in our loss of time, but much more in the fact that the scientist of today
is forced to subordinate himself to some counter-intellectual patterns of reports
and planning, forcing him indeed to accept the professional dishonesty. The most
absurd demand he faces is to present the program (and the time-table) of his future
discoveries. Such plans can bring the best results if they fail \dots The excursion of 
Christopher Columbus [did not] accomplish his original plan to discover the shortest 
way to India. The only thing discovered by CC was an obstacle, on which we live today!

\medskip 

New forms of business appear: the enterprises which help the scientists to formulate
their grant requests in terms convincing for the bureaucrats. (The corruptive
consequences are not difficult to guess!)

\bigskip

{\bf Four Laws of Bureaucracy}:
\smallskip

I. All attempts of the state administrations to improve the scientific work by
bureaucratic projects, reports, etc. will be reduced to zero by the social organism
– though not gratis: the price is an enormous increase of socially
useless work.

\smallskip
II. What is the source of the incredible facility of public administrations in multiplying
endlessly the prescriptions, formalities and obligatory documents?
The reason is that the bureaucrats do not perform the bureaucratic work:
they leave it to their victims.

\smallskip
III. In the bureaucratic environment the problems of little importance are always
infinitely more urgent than the truly important ones. This is why 
{\it thou will never do anything important}.

\smallskip
IV. The knowledge of the four bureaucracy laws won't help you in anything.
}

\end{document}